\documentclass[prd,letterpaper,twocolumn,preprintnumbers,nofootinbib]{revtex4}

\usepackage{amsmath,amssymb}
\usepackage{graphicx}
\usepackage{units}
\usepackage[hyperfootnotes=false,colorlinks,citecolor=blue]{hyperref}

\newcommand{\beq}{\begin{equation}}
\newcommand{\eeq}{\end{equation}}
\newcommand{\bea}{\begin{eqnarray}}
\newcommand{\ena}{\end{eqnarray}}
\newcommand{\dd}{{\rm d}}

\def \L {\mathcal{L}} 

\def \epsilon {\varepsilon} 

\def \vec#1{{\boldsymbol{#1}}}
\newcommand{\hc}{\ensuremath{\text{h.c.}}}
\newcommand{\BR}{\ensuremath{\text{BR}}}

\allowdisplaybreaks

\begin{document}

\title{Explaining lepton-flavor non-universality and self-interacting dark matter with $L_\mu-L_\tau$}

\author{Julian Heeck}
\email{heeck@virginia.edu}
\affiliation{Department of Physics, University of Virginia,
Charlottesville, Virginia 22904-4714, USA}

\author{Anil Thapa}
\email{wtd8kz@virginia.edu}
\affiliation{Department of Physics, University of Virginia,
Charlottesville, Virginia 22904-4714, USA}

\hypersetup{
pdftitle={Explaining lepton-flavor non-universality and self-interacting dark matter with Lmu - Ltau},   
pdfauthor={Julian Heeck, Anil Thapa}
}


\begin{abstract}
Experimental hints for lepton-flavor universality violation in the muon's magnetic moment as well as neutral- and charged-current $B$-meson decays require Standard-Model extensions by particles such as leptoquarks that generically lead to unacceptably fast rates of charged lepton flavor violation and proton decay. We propose a model based on a gauged $U(1)_{L_\mu-L_\tau}$ that eliminates all these unwanted decays by symmetry rather than finetuning and efficiently explains $(g-2)_\mu$, $R_{K^{(*)}}$, $R_{D^{(*)}}$, and neutrino masses. The $U(1)_{L_\mu-L_\tau}$ furthermore acts as a stabilizing symmetry for dark matter and the light $Z'$ gauge boson mediates velocity-dependent dark-matter self-interactions that resolve the small-scale structure problems. Lastly, even the Hubble tension can be ameliorated via the light $Z'$ contribution to the relativistic degrees of freedom.
\end{abstract}

\maketitle


\section{Introduction}

The Standard Model (SM) of particle physics is an immensely successful description of nature at the most fundamental level but necessarily needs to be extended by additional particles in order to accommodate the now-established fact that neutrinos are massive and mix with each other~\cite{Mohapatra:2005wg}.
In addition, hints for lepton-flavor universality violation (LFUV) beyond the SM have accumulated over the last years~\cite{Crivellin:2021sff}, notably in the following observables:
\begin{enumerate}
\item \textit{Muon's magnetic moment:} An approximately $3.5\sigma$ deviation of $(g-2)_\mu$ from the theoretically predicted SM value~\cite{Aoyama:2020ynm} was measured almost two decades ago at Brookhaven~\cite{Bennett:2006fi}. A recent measurement at Fermilab has confirmed this deviation and increased the significance to $4.2\sigma$~\cite{Muong-2:2021ojo}, to be improved even further in the near future. The theoretical prediction is under scrutiny as well, with some lattice-QCD results yielding values closer to the experimental one~\cite{Borsanyi:2020mff}.
Most lepton-flavor \emph{universal} new-physics models such as the dark photon cannot explain this $(g-2)_\mu$ deviation~\cite{Athron:2021iuf}, which therefore requires LFUV.\footnote{See, however, Refs.~\cite{Hiller:2019mou,Hiller:2020fbu} for flavor universal models.} Generically, such LFUV models lead to unacceptably fast rates for lepton flavor violation, which needs to be eliminated via finetuning or a symmetry~\cite{Isidori:2021gqe}.
\item \textit{Neutral-current $B$-meson decays:}  $B$-meson decays such as $B\to K^{(*)}\mu^+\mu^-$ exhibit anomalous behavior compared to the SM expectation, most importantly in the theoretically clean ratios
\begin{align}
R_{K^{(*)}} \equiv \frac{\Gamma (B\to K^{(*)}\mu^+\mu^-)}{\Gamma (B\to K^{(*)}e^+e^-)} \,,
\label{eq:RK}
\end{align}
which have been measured~\cite{Aaij:2014ora,Aaij:2017vbb,Aaij:2019wad,LHCb:2021trn} to lie several standard deviations below the SM predictions of $R_{K^{(*)}}^\text{SM}\simeq 1$~\cite{Bobeth:2007dw}.
In fact, global fits to all $b\to s \mu^+\mu^-$ data prefer the addition of one or two new-physics operators to the SM by more than $5\sigma$~\cite{Capdevila:2017bsm,Aebischer:2019mlg}, driven by $R_{K^{(*)}}$ and angular observables~\cite{LHCb:2020lmf} in these $B$ decays. The measured $R_{K^{(*)}}  < 1$ indicate LFUV, which can be realized at tree level through neutral vector bosons or leptoquarks.
\item \textit{Charged-current $B$-meson decays:}  
In addition to the LFUV observed in neutral-current $B$-meson decays, similar hints reside in the charged-current sector in measurements of the ratio
\begin{align}
R_{D^{(*)}} \equiv \frac{\Gamma (B\to D^{(*)}\tau\nu)}{\Gamma (B\to D^{(*)}\ell\nu)} \,,
\label{eq:RD}
\end{align}
where $\ell=e$ or $\mu$.
The experimental measurements of $R_{D^{(*)}}$~\cite{BaBar:2012obs, BaBar:2013mob, Belle:2015qfa, LHCb:2015gmp, Belle:2016dyj, Belle:2016kgw, LHCb:2017smo} differ from the SM predictions $R_{D}^\text{SM} = 0.299\pm 0.003$ and $R_{D^{*}}^\text{SM} = 0.258\pm 0.005$~\cite{HFLAV:2019otj} by $\sim 3\sigma$. A related observable, $R_{J/\psi}$~\cite{LHCb:2017vlu}, shows a mild deviation of $1.7 \sigma$. However, the current experimental uncertainty on $R_{J/\psi}$ is large and any new-physics scenario that explains the $R_{D^{(*)}}$ anomaly automatically resolves the $R_{J/\psi}$ anomaly as well. 
Unlike the neutral-current anomaly, a significant new-physics contribution to $R_{D^{(*)}}$ requires rather light new particles, close to TeV, leptoquarks being the most prominent viable example.
\end{enumerate}
The above three flavor anomalies hint at new physics in the muon and tauon sector, and there is no shortage of models that can explain one or even all of them.
The masses of the new particles are typically required to be around the TeV scale and lead to unacceptably large rates for lepton flavor or even baryon number violation unless the relevant Yukawa couplings are finetuned to be minuscule. In absence of any such signals it is clearly desirable to find (flavor) symmetries that would simply forbid the unwanted couplings~\cite{Hambye:2017qix,Davighi:2020qqa,Greljo:2021xmg,Greljo:2021npi,Davighi:2022qgb}.
The anomaly-free $U(1)_{L_\mu-L_\tau}$~\cite{He:1990pn,Foot:1990mn,He:1991qd,Heeck:2011wj} symmetry emerges as an ideal candidate for this task because it singles out muons and tauons, precisely the lepton flavors that exhibit LFUV anomalies.
As a gauge symmetry, it has already been employed in the past to resolve either the $(g-2)_\mu$ anomaly~\cite{Gninenko:2001hx,Baek:2001kca,Carone:2013uh,Altmannshofer:2014pba} or the neutral-current $B$-meson discrepancies~\cite{Altmannshofer:2014cfa,Crivellin:2015mga,Crivellin:2015lwa,Altmannshofer:2016oaq,Altmannshofer:2016jzy,Alguero:2022est} depending on the $Z'$ mass $m_{Z'}$ and its gauge coupling $g'$.
In connection with leptoquarks, it has been used to resolve $R_{K^{(*)}}$ without dangerous proton decays~\cite{Davighi:2020qqa,Greljo:2021xmg,Greljo:2021npi}, even together with $R_{D^{(*)}}$~\cite{Greljo:2021xmg}.
The same $L_\mu-L_\tau$ symmetry can be used to explain $R_{K^{(*)}}$ and $R_{D^{(*)}}$ without worrying about lepton flavor and baryon number violation~\cite{Greljo:2021xmg}.

Finally, one of the biggest shortcomings of the SM is the lack of a dark matter (DM) candidate. The existence of DM is well established across many scales and observations, so far exclusively through its gravitational interactions~\cite{Bertone:2004pz}. A potential first hint for non-gravitational DM interactions comes from structure-formation data~\cite{Kaplinghat:2015aga,Tulin:2017ara,Bullock:2017xww,Sagunski:2020spe} -- the core--cusp problem -- which not only prefers rather large DM--DM self-interactions but even a particular velocity dependence of said cross section that has become increasingly difficult to realize in new-physics models due to competing constraints~\cite{Bringmann:2016din,Hambye:2019tjt}.

Here, we present a model based on a gauged $U(1)_{L_\mu-L_\tau}$ that generates viable neutrino masses, resolves all three LFUV anomalies, and simultaneously provides DM with large velocity-dependent self-interactions that can explain the small-scale structure puzzle.
We give a quick overview of the model in Sec.~\ref{sec:model}, followed by a detailed discussion of neutrino masses in this model in Sec.~\ref{sec:neutrinos}. We discuss the explanations of the neutral- and charged-current $B$-meson anomalies in Sec.~\ref{sec:neutral} and Sec.~\ref{sec:charged}, respectively, and that of the muon's magnetic moment in Sec.~\ref{sec:muon}.
The details of our dark matter sector are presented in Sec.~\ref{sec:dm} and we summarize our findings in Sec.~\ref{sec:conclusion}.

\section{Model and Overview}
\label{sec:model}

The model at hand is very simple and individual parts have already been discussed in the literature. We extend the SM by three right-handed neutrinos $N_R$ and promote the anomaly-free global symmetry $U(1)_{L_\mu-L_\tau}$ to a gauge symmetry~\cite{He:1990pn,Foot:1990mn,He:1991qd,Heeck:2011wj}.
In order to generate viable neutrino mixing angles, we break the $U(1)_{L_\mu-L_\tau}$ by an SM-singlet scalar $\phi_1$ with $U(1)_{L_\mu-L_\tau}$ charge $1$. Overall, this generates a low-energy Majorana mass matrix for the active neutrinos via seesaw.

We furthermore introduce two leptoquarks (LQs) to the model, one $S_3 \sim (\overline{\vec{3}},\vec{3},1/3)$ with $U(1)_{L_\mu-L_\tau}$ charge $-1$ and one $S_1\sim (\overline{\vec{3}},\vec{1},1/3)$ with $U(1)_{L_\mu-L_\tau}$ charge $+1$, using the notation of Ref.~\cite{Dorsner:2016wpm}.
The $U(1)_{L_\mu-L_\tau}$ charge assignments of the LQs enforce a coupling of $S_3$ exclusively to muonic leptons and of $S_1$ exclusively to tauonic leptons. This automatically eliminates any and all charged lepton flavor violation mediated by the LQ~\cite{Hambye:2017qix,Greljo:2021xmg}. Even more importantly, the $U(1)_{L_\mu-L_\tau}$ charges forbid the di-quark couplings of $S_{1,3}$ that would otherwise lead to fast proton decay~\cite{Hambye:2017qix}. The surviving LQ Yukawa couplings are precisely those that can resolve $R_{K^{(*)}}$ and $R_{D^{(*)}}$~\cite{Greljo:2021xmg}, as shown many times in the literature~\cite{Angelescu:2021lln}.\footnote{\label{footnoteR2}Instead of $S_1$, an $R_2\sim (\vec{3},\vec{2},7/6)$ leptoquark with $U(1)_{L_\mu-L_\tau}$ charge $+1$ would work equally well here to explain $R_{D^{(*)}}$~\cite{Crivellin:2017zlb,Crivellin:2019dwb,Angelescu:2021lln}. } 

The $U(1)_{L_\mu-L_\tau}$ breaking necessary for viable neutrino mixing renders the gauge boson $Z'$ massive, although the mass is not determined by the neutrino sector. We will assume the $Z'$ mass to lie between $\sim\unit[5]{MeV}$ and $2m_\mu \simeq \unit[210]{MeV}$, where it can consistently resolve the $(g-2)_\mu$ anomaly for couplings $g'\sim 10^{-3}$. Higher $Z'$ masses are excluded by direct searches using the $Z'\to \mu^+\mu^-$ decay channel, forcing the $Z'$ to live in the kinematical region where it can only decay into neutrinos.

Finally, we also introduce new particles $\chi$ charged under $U(1)_{L_\mu-L_\tau}$ in such a way that they are stable, leading to DM. The light $Z'$ then allows us to generate large DM--DM self-interactions with just the right velocity dependence~\cite{Tulin:2017ara,Garani:2019fpa} to explain the small-scale structure issues while remaining safe from cosmic-microwave-background constraints~\cite{Bringmann:2016din,Hambye:2019tjt}. It is remarkable that the $U(1)_{L_\mu-L_\tau}$ gauge boson in the $(g-2)_\mu$-preferred parameter space has just the right properties to mediate these DM self-interactions consistently; most, if not all, other $Z'$ models would fail here.

Overall, the $U(1)_{L_\mu-L_\tau}$ symmetry has several useful features that we employ here:
\begin{enumerate}
\item Since the $Z'$ only couples to second and third generation leptons, the constraints are weak and allow for a rather light $Z'$ that can explain $(g-2)_\mu$. 
\item Lepton-flavor charged LQs enforce lepton-flavor non-universality and automatically eliminate the otherwise dangerous proton and lepton flavor violating decays.
\item Charging DM under the $U(1)_{L_\mu-L_\tau}$ eliminates the need for an ad-hoc discrete stabilizing symmetry and opens up a viable DM freeze-out channel. Furthermore, the light $Z'$ can mediate a large velocity-dependent DM self-interaction that resolves small-scale problems.
\end{enumerate}
Individual aspects could work with other $U(1)'$ symmetries~\cite{Wang:2021uqz,Davighi:2022qgb}, but only $U(1)_{L_\mu-L_\tau}$ has the right properties  to resolve and connect all of these anomalies simultaneously and without fine-tuning. This is a simple model with few new particles (collected in Tab.~\ref{tab:charges}) and parameters that consistently accommodates the most significant failures of the SM.

\begin{table}[t]
\renewcommand{\baselinestretch}{1.3}\normalsize 
\centering
\begin{tabular}[t]{ c c | c c}
Fermions & & Bosons & \\
\hline                               
$N_{R,e}$ & \quad$\left(\vec{1},\vec{1},0,0\right)$ & $\phi_1$ & \quad$\left(\vec{1},\vec{1},0,1\right)$\\
$N_{R,\mu}$ & \quad$\left(\vec{1},\vec{1},0,1\right)$ & $S_3$ & \quad$\left(\overline{\vec{3}},\vec{3},\tfrac13,-1\right)$\\
$N_{R,\tau}$ & \quad$\left(\vec{1},\vec{1},0,-1\right)$ & $S_1$ & \quad$\left(\overline{\vec{3}},\vec{1},\tfrac13,+1\right)$\\ 
\end{tabular}
\begin{tabular}[t]{ c c  c c}
\  \   $\chi$ \qquad &\  \qquad$\left(\vec{1},\vec{1},0,q\right)$ \  or &\   $\chi$ & \ \qquad$\left(\vec{1},\vec{1},0,q\right)$ 
\end{tabular}
\renewcommand{\baselinestretch}{1.0}\normalsize 
\caption{\label{tab:charges}
$SU(3)_C \times SU(2)_L \times U(1)_Y \times U(1)_{L_\mu-L_\tau}$ representations of non-SM particles. The leptoquark notation $S_{1,3}$ follows Ref.~\cite{Dorsner:2016wpm}.
$\chi$ is our dark-matter candidate, either a fermion or boson.
}
\end{table}

\section{Neutrino Masses}
\label{sec:neutrinos}

Neutrino masses in the model are induced at tree level via the type-I seesaw mechanism \cite{Minkowski:1977sc, Mohapatra:1979ia, Yanagida:1980xy, Gell-Mann:1979vob} as
\beq
 m_\nu \simeq - m_D\ m_R^{-1}\ m_D^T \, ,
 \label{eq:nu}
\eeq
where $m_D = v/\sqrt{2}\ {\rm Diag} (\lambda_e, \lambda_\mu, \lambda_\tau)$ is the diagonal Dirac mass matrix, proportional to the electroweak breaking scale $v = 1/\sqrt{\sqrt{2} G_F}\simeq \unit[246.2]{GeV}$, and $m_{R}$ the right-handed Majorana neutrino mass matrix, given by \cite{Araki:2012ip}
\beq
m_{R}=\begin{pmatrix}
M_{1} & a_{12}\langle\phi_{1}\rangle & a_{13}\langle\phi_{1}\rangle \\
a_{12}\langle\phi_{1}\rangle & 0 & M_{2} \\
a_{13}\langle\phi_{1}\rangle & M_{2} & 0 
\end{pmatrix}.
\label{eq:mR}
\eeq
Here, $M_{1,2}$ are the bare $U(1)_{L_\mu-L_\tau}$-symmetric mass term for $N_{R,e} N_{R,e}$ and $N_{R,\mu} N_{R,\tau}$ and $a_{ij}$ are the Yukawa couplings associated with the $a_{ij} \phi_1 N_{Ri} N_{Rj}$ interactions, which lead to $L_\mu-L_\tau$ breaking terms in $m_R$. 
The surviving texture zeros $(m_R)_{22}=(m_R)_{33}=0$ lead to two vanishing \emph{minors}~\cite{Lavoura:2004tu,Lashin:2007dm,Araki:2012ip,Crivellin:2015lwa,Asai:2018ocx} in the neutrino mass matrix, $(m_\nu^{-1})_{22} = (m_\nu^{-1})_{33} = 0$, which lead to four relations between the neutrino mixing parameters. Using $\theta_{12}$, $\theta_{23}$, $\theta_{13}$, $\Delta m_{13}^2$, and $\Delta m_{12}^2$ as input parameters, we can predict the three unknown CP phases and the overall neutrino mass scale. Our vanishing minors are only compatible with normal hierarchy, and using the best-fit values from Ref.~\cite{Esteban:2020cvm}, we predict
\begin{align}
\sum_j m_j = \unit[0.17]{eV}\,, &&
m_{\beta\beta} = \unit[0.036]{eV}\,, &&
\delta= 246^\circ\text{ or } 114^\circ,
\end{align}
for the sum of neutrino masses, the effective neutrino mass for neutrinoless double-beta decay~\cite{Rodejohann:2011mu}, and the Dirac CP phase $\delta$, respectively.
While the latter two predictions are perfectly compatible with observations, the sum of neutrino masses has an effect on cosmology that leads to rather strong constraints, ranging from $\sum_j m_j < \unit[0.16]{eV}$ to $\sum_j m_j < \unit[0.12]{eV}$ at $95\%$\,C.L., depending on the combination of data sets~\cite{Vagnozzi:2017ovm,Planck:2018vyg,deSalas:2020pgw} and underlying cosmological model~\cite{Vagnozzi:2018jhn}. The best-fit prediction of our two vanishing minors is hence comfortably excluded by cosmological data by now, as already pointed out in Ref.~\cite{Asai:2018ocx}. In fact, even varying the neutrino oscillation parameters within their $2\sigma$ ranges from Ref.~\cite{Esteban:2020cvm}  can only push $\sum_j m_j $ down to $\unit[0.124]{eV}$,\footnote{In this region, we need $\theta_{23}\simeq 51^\circ$ and $\delta \simeq 230^\circ$ or $130^\circ$.} which is still not enough to satisfy the strongest cosmological bounds, thus excluding our texture zeros at the $2\sigma $ level. 
If the cosmological constraint $\sum_j m_j < \unit[0.12]{eV}$ is solidified -- which, at the very least, requires a better understanding of the Hubble tension~\cite{DiValentino:2021izs,Dainotti:2021pqg,Dainotti:2022bzg} -- the two vanishing minors are excluded and our model needs to be amended by a second scalar $\phi_2 \sim (\vec{1},\vec{1},0,2)$ with a vacuum expectation value that generates the entries $(m_R)_{22}$ and $(m_R)_{33}$, filling the vanishing minors. That model can accommodate all neutrino data without any predictions and would have no impact on the rest of this article.

As we will see later, the $\phi_1$ vacuum expectation value $\langle \phi_1\rangle = m_{Z'}/g'$ is fixed to be $\mathcal{O}(\unit[100]{GeV})$ in order to resolve the muon's magnetic moment anomaly. The entries in $m_R$ are then expected to be of similar order, predicting right-handed Majorana neutrinos at the electroweak scale.
They can, in particular, be light enough to be kinematically accessible in $S_1$ LQ decays $S_1\to d_j N_R$, followed by $N_R$ decays.

\section{Neutral-Current B-Meson Anomalies}
\label{sec:neutral}

The $S_3$ LQ from Tab.~\ref{tab:charges} couples exclusively to muonic leptons via the simple Lagrangian
\begin{align}
\L_{S_3} &= y_j \overline{Q}^c_{j} S_3 P_L L_{\mu} +\hc  ,
\label{eq:S3lag}
\end{align}
whose simple lepton-flavor structure automatically eliminates lepton flavor violation. Even more importantly, the $L_\mu-L_\tau$ charge of $S_3$ forbids the otherwise allowed di-quark coupling $Q Q S_3^*$ that would lead to disastrous proton decay rates~\cite{Hambye:2017qix,Davighi:2020qqa,Greljo:2021xmg,Greljo:2021npi}. The muon-flavored $S_3$ LQ thus naturally evades all constraints from lepton-flavor~\cite{Crivellin:2019dwb} and baryon-number violating decays and can easily have a mass close to the electroweak scale. $S_3$ contributes at tree level to the decay $b \to s \mu^+ \mu^-$ that appears to deviate from the SM prediction~\cite{Aebischer:2019mlg}. Assuming $S_3$ to be heavy, we can integrate it out to obtain the purely left-handed Wilson coefficients~\cite{Angelescu:2018tyl}
\beq
C_9^{\mu \mu} = - C_{10}^{\mu \mu} = \frac{\pi v^2}{V_{tb} V_{ts}^* \alpha_\text{EM}} \frac{y_{b} y_{s}^*}{m_{S_3}^2} \, ,
\label{eq:C9}
\eeq
with the electromagnetic fine-structure constant $\alpha_\text{EM}$ and the Cabibbo--Kobayashi--Maskawa quark-mixing matrix $V$ arising as a convenient normalization of the new-physics operator strength.
The required best-fit value of the real part of these Wilson coefficients at the scale $\mu= m_b$ is $-0.41\pm 0.09$~\cite{Aebischer:2019mlg, Angelescu:2021lln}  which accounts for the LFUV ratios $(R_{K}, R_{K^*})$ as well as the discrepancy in the angular observable $P_5'$ in the $B \to K^* \mu \mu$ decay \cite{LHCb:2015svh, CMS:2015bcy, ATLAS:2018gqc}.

\begin{figure}
    \centering
    \includegraphics[width=0.49\textwidth]{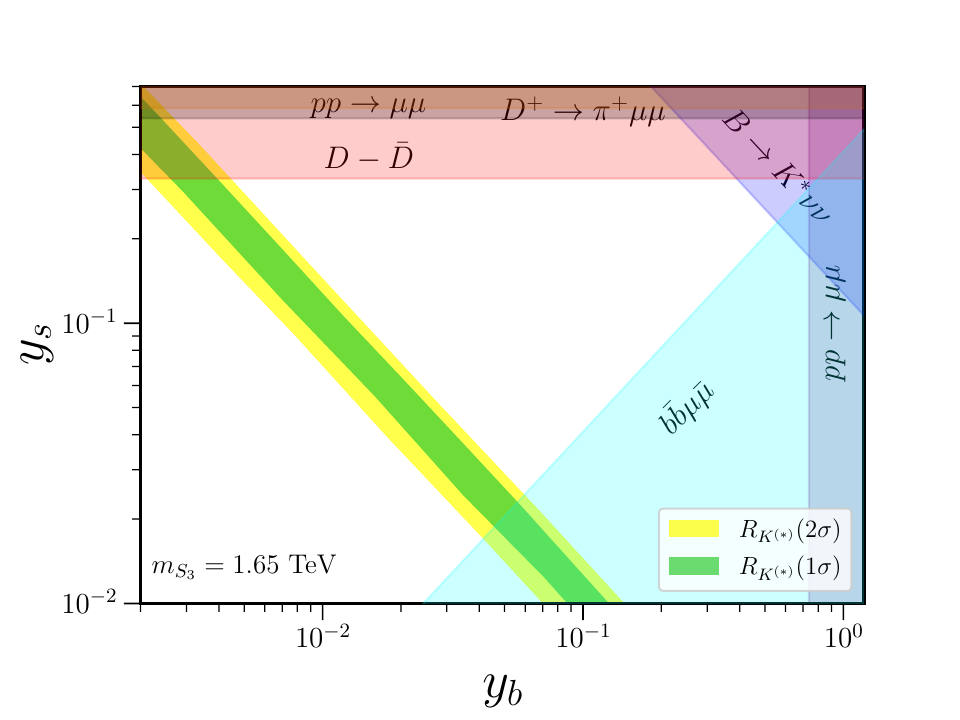}
    \caption{$1\sigma$ (green) and $2\sigma$ (yellow) allowed range for $R_{K^{(*)}}$ while accounting for $\BR(B_s \to \mu \mu)$ in the relevant Yukawa coupling plane, with $S_3$ LQ mass fixed at \unit[1.3]{TeV}.  Cyan and vertical purple shaded regions are exclusion limits obtained from LQ pair production in $pp \to b \bar{b} \mu \bar{\mu}$~\cite{ATLAS:2020dsk} and most recent LHC searches in the high-$p_T$ bins of $pp \to \mu \mu$ at $\sqrt{s} = \unit[13]{TeV}$ with $\unit[140]{fb^{-1}}$~\cite{ATLAS:2020zms, CMS:2019tbu}. The blue, red, and black regions correspond to the exclusion from $B \to K^* \nu \nu$ \cite{Buras:2014fpa}, $D-\bar{D}$ mixing \cite{Crivellin:2019qnh, Bazavov:2017lyh}, and $D^+ \to \pi^+ \mu \mu$ decay \cite{Babu:1987xe, Babu:2019mfe}.      
     }
    \label{fig:RK}
\end{figure}

The best-fit value $C_9^{\mu \mu} = - C_{10}^{\mu \mu} =-0.41$ corresponds to a LQ mass $m_{S_3} \sim \unit[40]{TeV}\times\sqrt{|y_{b} y_{s}|}$, 
 which can easily be large enough to evade direct-search constraints from the LHC.
 Lowering the $S_3$ mass and the $y_{b,s}$ couplings -- so as to keep the contribution to Eq.~\eqref{eq:C9} fixed -- brings about complementary constraints on the model, including $D$--$\bar{D}$ mixing and pair production at the LHC. Using current data, we established a lower bound $m_{S_3} > \unit[1.45]{TeV}$ that can still resolve the neutral-current $B$-meson anomalies. An example of the constraints for a low-mass LQ scenario $m_{S_3} = \unit[1.65]{TeV}$ is shown in Fig.~\ref{fig:RK}.

To resolve the neutral-current $B$-physics anomalies, we only require the couplings $y_s$ and $y_b$ in Eq.~\eqref{eq:S3lag}. The first-generation quark coupling $y_d$ is, of course, also allowed, and constrained  through leptonic Kaon decays~\cite{Mandal:2019gff} such as $K_{S,L}^0 \to \mu^+ \mu^-$, $K^+ \to \pi^+ \mu^+ \mu^- (\pi^+ \nu \bar{\nu})$, and $K_{S,L}^0 \to \pi^0 \mu^+ \mu^-$. Among these, the most stringent constraint comes from $K_L^0 \to \mu^+ \mu^-$, which reads as $|\Re (y_{d}^* y_{s})| < 1.17 \times 10^{-5}\ (m_{S_3}/{\rm TeV})^2$, and $K^+ \to\pi^+ \nu \bar{\nu}$, which gives $|\Re (y_{d} y_{s}^*)| \sim [-3.7,8.3] \times 10^{-4}\ (m_{S_3}/{\rm TeV})^2$. Moreover, $S_3$ contributes to kaon mixing via a box diagram, yielding the constraint $|y_{d} y_{s}^*| < 0.013\ (m_{S_3}/{\rm TeV})$.  
Overall, the $y_d$ coupling is roughly constrained to lie below $10^{-2}$ for $m_{S_3}\sim\unit[30]{TeV}$ and $10^{-4}$ for $m_{S_3}\sim\unit[1]{TeV}$,
hardly in the fine-tuned regime.

\section{Charged-Current B-Meson Anomalies}
\label{sec:charged}

To address the \emph{charged}-current $B$-meson anomalies, we employ the $S_1$ LQ\textsuperscript{\ref{footnoteR2}} from Tab.~\ref{tab:charges} with Lagrangian
\begin{align}
\begin{split}
\L_{S_1} &=  z^L_j \overline{Q}^c_{j} S_1 P_L  L_{\tau}+ z^R_j \overline{u}^c_{j} S_1  P_R \tau\\
&\quad+ z^N_j \overline{d}^c_{j} S_1 P_R  N_{R,\tau}  +\hc 
\end{split}
\label{eq:S1lag}
\end{align}
Just like for the $S_3$ LQ above, the $L_\mu-L_\tau$ charge of $S_1$ automatically eliminates any and all lepton-flavor and baryon-number violation~\cite{Greljo:2021xmg}. Without this symmetry, dozens of $S_1$ Yukawa couplings would have to be tuned to minuscule values in order to suppress unwanted proton, muon, and tauon decays.

The tau-flavored $S_1$ LQ contributes at tree level to the decay $b \to c \tau \nu_\tau$ that shows disagreement with the SM prediction. There are no new-physics contributions from the electron and muon sector as $S_1$ exclusively couples to tauonic leptons due to our $L_\mu-L_\tau$ symmetry. Integrating out the $S_1$ field  gives the following relevant Wilson coefficients of the effective Hamiltonian~\cite{Jung:2018lfu}:
\begin{align}
    C_s^\tau &= -4 C_T^\tau = -\frac{v^2}{4 V_{cb}} \frac{z_{b}^L z_{c}^{R*}}{m_{S_1}^2} \, , \label{eq:gs}\\
    C_V^\tau &=   \frac{v^2}{4 V_{cb}} \frac{z_{b}^L (Vz^{L*})_{c}}{m_{S_1}^2}\ \, .
    \label{eq:gV}
\end{align}
The contribution of these Wilson coefficients to the LFUV ratios $R_{D^{(*)}}$  with $\nu_\tau$ in the final state are approximately given by~\cite{Blanke:2018yud}
 \begin{align}
     R_D &\simeq R_D^{\rm SM} \Big( |1+C_V^\tau|^2 + 1.54\, \Re [C_s^{\tau*} (1+ C_V^\tau)] + 1.09 |C_s^\tau|^2 \nonumber\\
     &~~~~~~~ + 1.04\, \Re [C_T^{\tau*} (1+ C_V^\tau)] + 0.75 |C_T^\tau|^2 \Big) \,,\\
     R_{D^*} &\simeq R_{D^*}^{\rm SM} \Big( |1+C_V^\tau|^2 - 0.13\, \Re [C_s^{\tau*} (1+ C_V^\tau)] + 0.05 |C_s^\tau|^2 \nonumber\\
     &~~~~~~~ -5.0\, \Re [C_T^{\tau*} (1+ C_V^\tau)] + 16.27 |C_T^\tau|^2 \Big) \,.
 \end{align}
 The above procedure neglects renormalization group running of the Wilson coefficients; numerically, we match the $S_1$ operators to coefficients in the SM Effective Field Theory at the scale  $\mu= m_{S_1}$, which are then run down to the $b$-quark scale  $\mu = m_b$ and matched to the effective Hamiltonian~\cite{Jung:2018lfu}  using the {\tt Flavio/Wilson} package~\cite{Straub:2018kue, Aebischer:2018bkb} (cf.~Ref.~\cite{Dorsner:2013tla, Gonzalez-Alonso:2017iyc}). 
These Wilson coefficients also lead to $B_c \to \tau \nu_\tau$ decay \cite{Alonso:2016oyd, Celis:2016azn, Li:2016vvp, Blanke:2018yud} with the associated branching ratio given as
\begin{equation}
    \BR (B_c \to \tau \nu_\tau) \simeq 0.023\ |1+ C_V^\tau - 4.3\ C_s^\tau|^2 \, ,
\end{equation}
where $C_s^\tau$ and $C_V^\tau$ are evaluated by running the Wilson coefficient to the $\mu = m_{B_c}$ scale \cite{Straub:2018kue, Gonzalez-Alonso:2017iyc}.  
Since this branching ratio has not been measured yet experimentally, we impose a conservative limit $\BR (B_c \to \tau \bar{\nu}) \lesssim 30 \%$ in our analysis~\cite{Alonso:2016oyd}.

\begin{figure}
    \centering
    \includegraphics[width=0.49\textwidth]{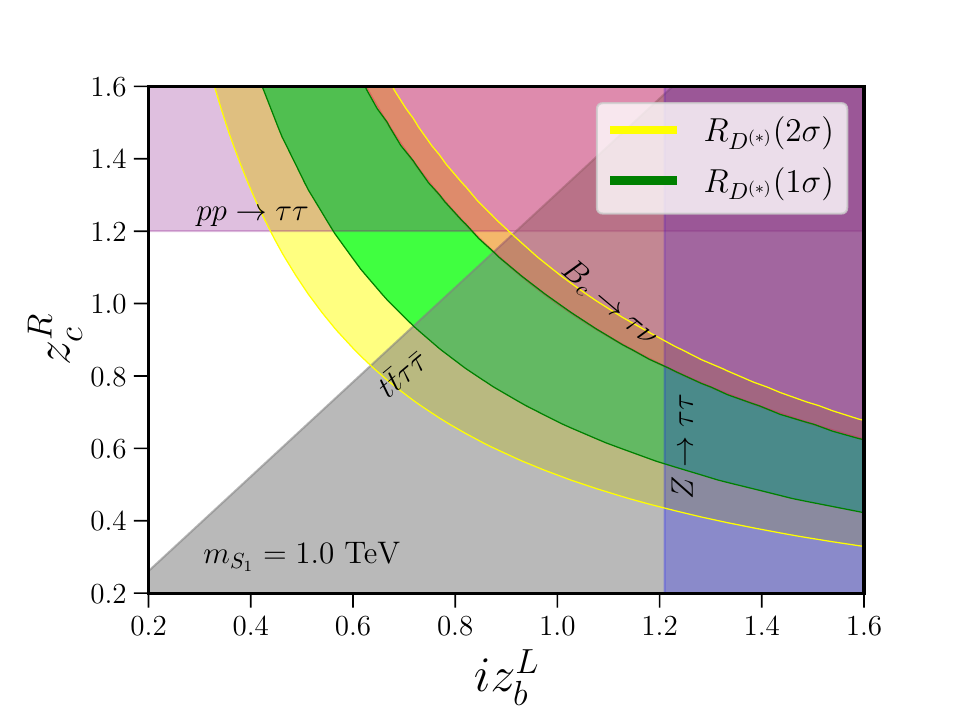}
    \includegraphics[width=0.49\textwidth]{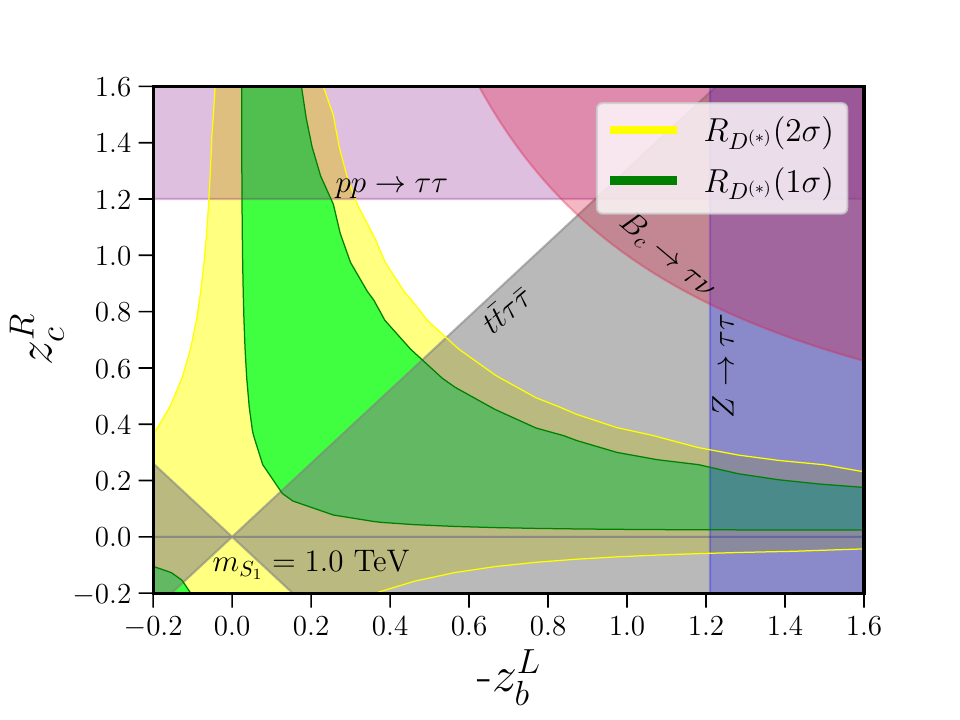}
    \caption{$1\sigma$ (green) and $2\sigma$ (yellow) allowed range for $R_{D^{(*)}}$ in the relevant Yukawa coupling plane, with the $S_1$ LQ mass fixed at $\unit[1.0]{TeV}$ for the case with $C_s^\tau = - 4 C_T^\tau \in i \Re$ (top) and $C_s^\tau = - 4 C_T^\tau \in \Re$ (bottom). The vertical blue band corresponds to the $Z\to \tau \tau$ constraint. The gray and purple band  represent exclusions from LQ pair production in $pp \to t \bar{t} \tau \bar{\tau}$ \cite{ATLAS:2021oiz} and from the most recent LHC searches in the high-$p_T$ bins of $pp \to \tau \tau$ at $\sqrt{s}= \unit[13]{TeV}$ with $\unit[140]{fb^{-1}}$ \cite{ATLAS:2020zms, CMS:2019tbu}, respectively. The red curved band corresponds to the exclusion from $B_c \to \tau \nu_\tau$ decay with the conservative limit on branching ratio $< 30 \%$~\cite{Alonso:2016oyd}. }
    \label{fig:RD}
\end{figure}

The two independent Wilson coefficients $C_s^\tau$ and $C_V^\tau$ generated by $S_1$ lead to a range of possible  $R_{D^{(*)}}$ explanations.
First, we discuss the possibility of accommodating $R_{D^{(*)}}$ purely from $C_s^\tau$ given in Eq.~\eqref{eq:gs}. This can be achieved by taking the coupling $z_{c}^L \ll 1$ such that the contribution from $C_V^\tau$ is negligible. Furthermore, we first consider the case with $C_s^\tau = - 4 C_T^\tau \in i \Re$ as shown in Fig.~\ref{fig:RD} (top). 
This scenario requires the $z_{t}^L \equiv z_{b}^L$ coupling to be of $\mathcal{O}(1)$, which modifies $Z$-boson decays to fermion pairs through one-loop radiative corrections mediated by $S_1$. Using the experimental results on the effective coupling obtained by the LEP collaboration \cite{ALEPH:2005ab}, we obtain $2\sigma$ limit $|z_{b}^L| \leq 1.21$ \cite{Arnan:2019olv} for the LQ mass of \unit[1.0]{TeV}, as represented by the purple region in Fig.~\ref{fig:RD}.

At the LHC, the $S_1$ LQ can be pair produced \cite{Diaz:2017lit, Dorsner:2018ynv} through $gg$ and $q\bar{q}$ fusion processes, or can be singly produced in association with charged leptons via $s$- and $t$- channel quark-gluon fusion processes. The single production of LQ becomes  relevant only for larger Yukawa couplings to the first and second generation quarks \cite{Buonocore:2020erb}. Thus, collider bounds from single-LQ production are less significant compared to QCD-driven LQ pair production. $S_1$ pairs primarily decay to $b\bar{b} \nu \nu$ \cite{CMS:2018bhq} and $t\bar{t}\tau \bar{\tau}$ \cite{ATLAS:2021oiz} (branching ratio $\beta \leq 0.5$) via the coupling $z_{b}^L$, and to $jj\tau\bar{\tau}$ through the coupling $z_{c}^R$. There are no dedicated searches for $jj\tau\bar{\tau}$ decays and the limit from $b\bar{b} \nu \nu$ is $m_{S_1} >\unit[1.0]{TeV}$ for $\beta =1$. The only important pair production bound is from $t\bar{t}\tau \bar{\tau}$, which is shown as gray region in Fig.~\ref{fig:RD}. Apart from the LQ-pair production bound, there are bounds on the couplings and mass on the LQ from the high-$p_T$ tails of $pp\to \ell \ell$ distributions \cite{Eboli:1987vb, Faroughy:2016osc, Schmaltz:2018nls, Greljo:2017vvb, Alves:2018krf, Angelescu:2020uug, Babu:2020hun}. The bound on $pp \to \tau \tau$ at $\sqrt{s}= \unit[14]{TeV}$ with $\unit[140]{fb^{-1}}$ for the Yukawa coupling is $z_{c}^R \leq 1.2$ \cite{Angelescu:2021lln} for a  \unit[1]{TeV} LQ mass as shown by pink region in Fig.~\ref{fig:RD}.

A similar analysis is performed considering the case with $C_s^\tau = - 4 C_T^\tau \in \Re$ \cite{Straub:2018kue}, still assuming $C_V^\tau$ to be negligible. The allowed region, along with different constraints, is shown in Fig.~\ref{fig:RD} (bottom). 
There is evidently ample room to accommodate $R_{D^{(*)}}$ while evading other constraints, typically requiring the LQ mass to be around TeV and the relevant couplings of order one.

Having neglected $C_V^\tau$ so far, let us consider the opposite scenario now where $C_V^\tau$ dominates, for example because $z^R\simeq 0$. The allowed $1 \sigma$ range to explain $R_{D^{(*)}}$ is $C_V^\tau = 0.07 \pm 0.02$ \cite{Angelescu:2021lln}. Note that the product of Yukawa couplings $z_{b}^L z_{c}^L$ suffer a strong constraint from $B \to K^{(*)} \nu \bar{\nu}$ decay at tree-level via $b \to s \nu_\tau \nu_\tau$ processes~\cite{Buras:2014fpa}: $z_{b}^L z_{s}^L <  0.076\ (0.087)\ (m_{S_1}/\text{TeV})^2$ obtained from $R_{K^*}^{\nu \bar{\nu}} < 2.7$ ($R_{K}^{\nu \bar{\nu}} < 3.9$) \cite{Belle:2017oht}. One can, however, obtain the best fit value for $C_V^\tau$ with just a third generation Yukawa coupling $z_{b}^L$ in conjunction with CKM mixing that corresponds to $|z_{b}^L| \simeq 2.15\ (m_{S_1}/{\rm TeV})$.
Finally, turning on both $C_s^\tau$ and $C_V^\tau$ opens up even more parameter space to explain $R_{D^{(*)}}$ that we will not dissect here.

Let us note that one of the peculiarities of our $S_1$ model here is the possible decay into right-handed neutrinos $N_R$. As we will see in the next section, the $L_\mu-L_\tau$ breaking vacuum expectation value $\langle \phi_1\rangle$ needs to be at the electroweak scale in order for the $Z'$ to resolve the $(g-2)_\mu$ anomaly. Together with the seesaw scenario described in Sec.~\ref{sec:neutrinos}, this implies right-handed neutrino masses below the TeV scale.
The $z^N$ coupling in Eq.~\eqref{eq:S1lag} then induces $S_1\to \bar{d}_j N_R$ with a potentially soft anti-quark if $N_R$ is not far below $m_{S_1}$. If the $S_1$ branching ratios into $N_R$ are large, all collider constraints will be appropriately diluted given the absence of searches for $S_1\to \bar{d}_j N_R$. This opens up the viable parameter space to explain $R_{D^{(*)}}$ even more.

\section{Muon's Magnetic Moment}
\label{sec:muon}

For the LQ explanations of the $B$-meson anomalies above it was not important whether the $U(1)_{L_\mu-L_\tau}$ is a global or gauged symmetry, as either one would have imposed the desired structure on the Yukawa couplings. In order to explain the $(g-2)_\mu$ anomaly, however, we require a \emph{gauged} symmetry. 
This is because our LQs cannot explain $(g-2)_\mu$ since $S_3$ is a chiral LQ and $S_1$ does not couple to muons at all in our model.
Gauging $U(1)_{L_\mu-L_\tau}$ introduces a $Z'$ that couples exclusively to muons, tauons, and their neutrinos, which in particular generates a one-loop contribution to the anomalous magnetic moment of the muon $a_\mu = (g-2)_\mu/2$ of~\cite{Baek:2001kca}
\begin{align}
    \Delta a_\mu = \frac{{g'}^2}{4\pi} \frac{1}{2\pi} \int_0^1\dd x \,\frac{2 m_\mu^2 x^2 (1-x)}{m_\mu^2 x^2 + m_{Z'}^2 (1-x)} \,.
\end{align}
This has long been a motivation for a $U(1)_{L_\mu-L_\tau}$ gauge symmetry and continues to be one of the simplest models that can explain the deviation in $(g-2)_\mu$. We require $\Delta a_\mu = (251\pm 59)\times 10^{-11}$~\cite{Muong-2:2021ojo}, which constitutes $4.2\sigma$ discrepancy from the SM prediction.

\begin{figure}[tb]
	\includegraphics[width=0.52\textwidth]{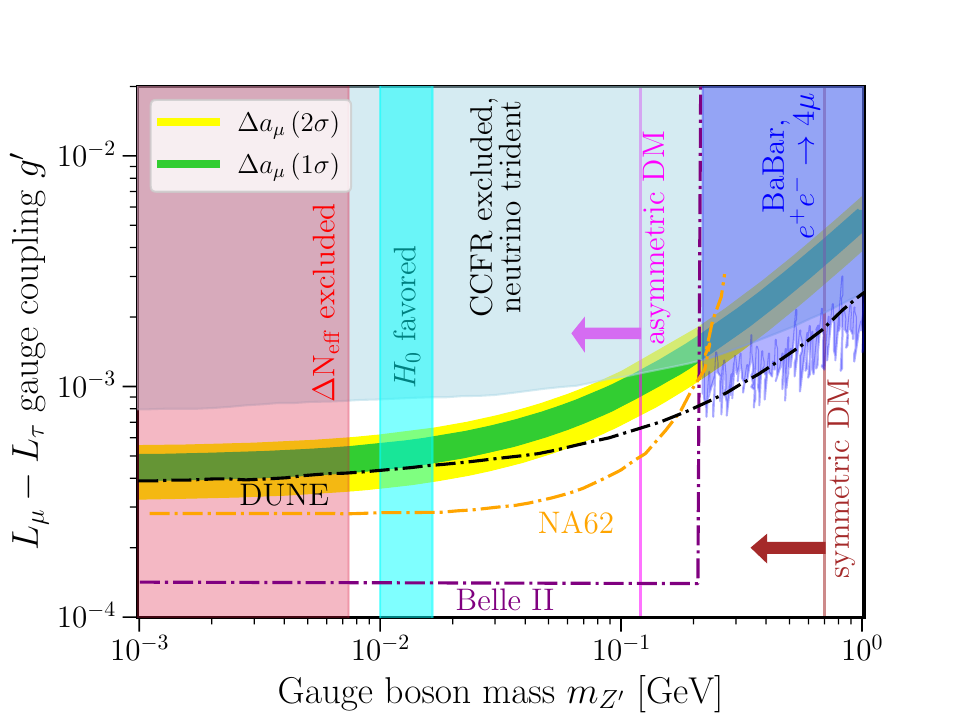}      
	\caption{
	Limits on the $U(1)_{L_\mu-L_\tau}$ gauge boson mass $m_{Z'}$ and coupling $g'$, assuming vanishing kinetic mixing. 
	The green (yellow) region can resolve the $(g-2)_\mu$ anomaly~\cite{Muong-2:2021ojo} at $1\sigma$ ($2\sigma$).
The  cyan region ameliorates the Hubble $H_0$ tension~\cite{Escudero:2019gzq}.
	The other shaded regions are excluded by $N_\text{eff}$~\cite{Kamada:2015era,Kamada:2018zxi}, BaBar~\cite{TheBABAR:2016rlg}, and neutrino trident production in CCFR~\cite{Mishra:1991bv,Altmannshofer:2014pba}.
	The dashed lines show the expected sensitivities of Belle-II (with $\unit[10]{ab^{-1}}$)~\cite{Jho:2019cxq}, DUNE~\cite{Altmannshofer:2019zhy,Ballett:2019xoj} and NA62 (in $K\to \mu +\text{inv}$)~\cite{Krnjaic:2019rsv}. Not shown are the sensitivities of M$^3$~\cite{Kahn:2018cqs} and NA64$\mu$~\cite{Gninenko:2014pea,Chen:2018vkr}. The vertical magenta (brown) solid line with arrow pointing to the left corresponds to the preferred region for asymmetric (symmetric) DM that solves the small-scale structure problems, see text for details. 
}
	\label{fig:Lmu-Ltau_limits}
\end{figure}

Other observables restrict the viable region where the $Z'$ can explain the anomaly to the mass region between a few MeV and $2 m_\mu$, see Fig.~\ref{fig:Lmu-Ltau_limits}. The upper bound comes from direct searches, which currently only sensitive to the $Z'$ mass region that kinematically allows for the decay $Z'\to\mu^+\mu^-$.
For $m_{Z'} < 2 m_\mu$, the $Z'$ can only decay into neutrinos, making direct searches difficult despite the low mass and rather large coupling. Nevertheless, the $(g-2)_\mu$-motivated region of parameter space will be thoroughly explored in several experiments in the near future and conclusively decide whether  $U(1)_{L_\mu-L_\tau}$ is the culprit for the muonic discrepancy.
The \emph{lower} bound on $m_{Z'}$ comes from the light-$Z'$ contribution to the radiation density of the universe, usually parametrized as $\Delta N_\text{eff}$, the effective number of additional neutrinos. Constraints on this quantity from Big Bang nucleosynthesis and the cosmic microwave background result in lower bounds on $m_{Z'}$ between $\unit[3]{MeV}$ and $\unit[10]{MeV}$, depending on the combination of datasets~\cite{Sabti:2019mhn}. In fact, since $N_\text{eff}$ measurements in the cosmic microwave background are partially degenerate with measurements of the Hubble constant $H_0$, it is possible to alleviate the current Hubble tension by allowing for a small non-SM contribution $\Delta N_\text{eff}\simeq 0.3$~\cite{Vagnozzi:2019ezj,DiValentino:2021izs}. In our model, this would imply a preferred $Z'$ mass around $\unit[10]{MeV}$~\cite{Escudero:2019gzq,Drees:2021rsg}, as indicated in Fig.~\ref{fig:Lmu-Ltau_limits}.

As we will see below, the required lightness of the $Z'$ in order to explain $(g-2)_\mu$ is not only a welcome ingredient for dark matter, but even \emph{required} to generate sufficiently large and velocity-dependent DM--DM self-interactions.

\section{Dark Matter}
\label{sec:dm}

In addition to explaining neutrino masses and LFUV, our $U(1)_{L_\mu-L_\tau}$ model can easily accommodate dark matter and even resolve small-scale structure problems through large DM--DM self-interactions, mediated by the light $Z'$. A welcome side effect is the stabilization of the DM particle by the $U(1)'$ rather than an ad-hoc discrete symmetry.

DM charged under $U(1)_{L_\mu-L_\tau}$ has been discussed for some time, typically as a weakly-interacting massive particle with thermal abundance~\cite{Cirelli:2008pk,Baek:2008nz,Foldenauer:2018zrz,Okada:2019sbb,Holst:2021lzm,Drees:2021rsg}. An appropriately chosen $L_\mu-L_\tau$ charge $q$ can render a new particle $\chi$ stable with $Z'$ connections to muons, tauons, and neutrinos.\footnote{Here we ignore kinetic mixing and scalar-DM couplings to the Higgs for simplicity, seeing as these come with free parameters.}
For most values of $q$, the full Lagrangian actually has an additional \emph{global} symmetry $U(1)_\chi$ that corresponds to conserved DM number. In analogy to the observed baryon number asymmetry one could assume that the cosmological history also led to a DM asymmetry~\cite{Zurek:2013wia} which results in a density $\Omega_\text{asym}$ proportional to the asymmetry.
Unless the $Z'$ couplings are tiny, the DM $\chi$ will also be produced thermally, leading to a \emph{symmetric} DM abundance~$\Omega_\text{sym}$. We will consider both options here but remain agnostic about the potential DM asymmetry.

Since our $Z'$ mass is forced to be in the sub-GeV range in order to resolve the $(g-2)_\mu$ anomaly (Fig.~\ref{fig:Lmu-Ltau_limits}), the relevant hierarchy in the DM sector is $m_{Z'}\ll m_\chi$. This not only opens up the annihilation channel $\bar{\chi}\chi\to Z' Z'$ but renders it dominant, with approximate cross section~\cite{Altmannshofer:2016jzy,Arcadi:2017kky}
\begin{align}
\langle \sigma v \rangle (\bar{\chi}\chi \to Z' Z') \simeq \xi \frac{(q g')^4}{8\pi m_\chi^2}\,,
\end{align}
where $\xi=1$ for complex-scalar DM and $\xi = 1/2$ for Dirac DM.
The competing annihilation channels $\bar{\chi}\chi\to$\,leptons via $s$-channel $Z'$ are suppressed by $1/q^2$, which is small in the region of interest here, as well as velocity suppressed in the scalar DM case.
For the thermal-relic case, we require $\Omega_\text{asym}h^2\ll \Omega_\text{sym}h^2 \simeq  0.12$~\cite{Planck:2018vyg}, which approximately translates to $q g' \simeq 0.02 \sqrt{m_\chi/\unit{GeV}}$. In the asymmetric DM case, the coupling $q g'$ needs to be even larger in order to suppress $\Omega_\text{sym}h^2 \ll 0.12$.

\begin{figure}
    \includegraphics[width=0.52\textwidth]{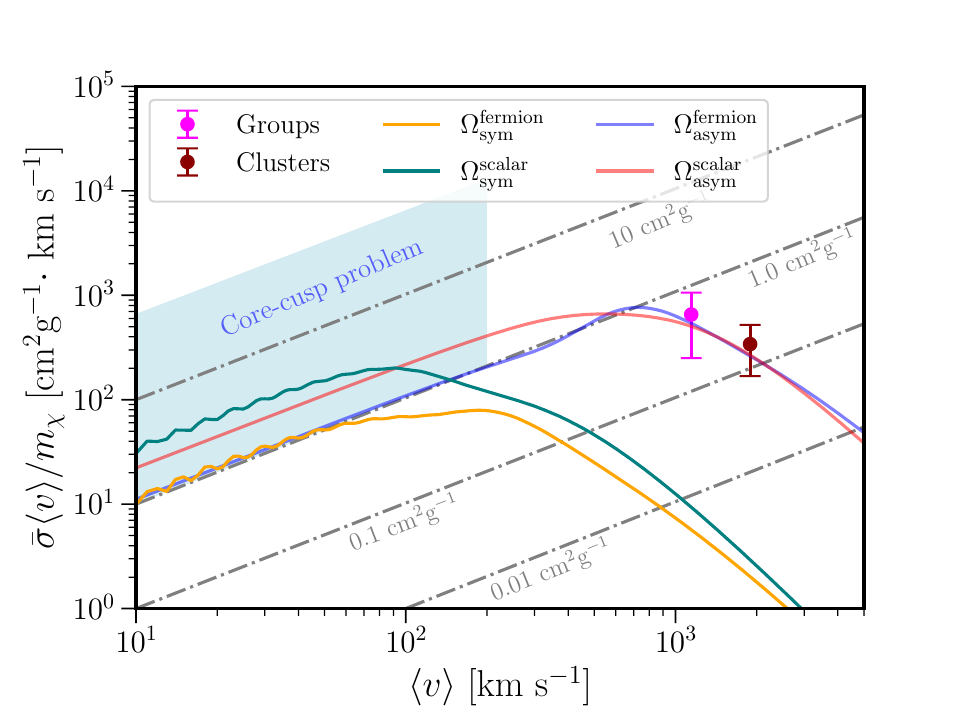}
    \caption{Red and blue (green and orange) lines correspond to average DM viscosity self-interaction cross sections $\bar\sigma$ for scalar and fermionic asymmetric (symmetric) DM. Here, $m_{Z'} = \unit[10]{MeV}$ with DM mass $m_\chi = 4\ \unit[(15)]{GeV}$ for asymmetric (symmetric) DM. The green and orange lines have $\Omega_\text{asym}h^2 \ll \Omega_\text{sym}h^2\simeq 0.12$, the red and blue lines have $\Omega_\text{asym}h^2 \simeq 100\,\Omega_\text{sym}h^2\simeq 0.12$.
    Diagonal dot-dashed lines are contours of constant $\bar\sigma/m_\chi$; data points and core--cusp problem region from Refs.~\cite{Kaplinghat:2015aga,Sagunski:2020spe}.
    The small wiggles at small $\langle v\rangle$ are an unphysical artifact of interpolation.}
    \label{fig:sig_V}
\end{figure}

For a given DM mass, the desired relic abundance can be fixed using the unknown $L_\mu-L_\tau$ charge $q$. Due to the dominance of the $\bar{\chi}\chi\to Z' Z'$ channel, the abundance only depends on the combination $q g'$, which is the same combination that enters the DM--DM self-interaction cross sections $\bar{\chi}\chi \to  \bar{\chi}\chi$ and $\chi\chi\to \chi\chi$, mediated by the $Z'$.\footnote{Using instead a $U(1)_{L_\mu-L_\tau}$-breaking scalar to mediate the DM scattering was discussed in Refs.~\cite{Kamada:2018zxi,Kamada:2020buc}.} For a fixed relic abundance, the DM self-interaction therefore only depends on the two masses $m_\chi$ and $m_{Z'}$.
The self-interaction cross sections through the $Z'$ Yukawa potential can be calculated analytically in parts of the parameter space~\cite{Feng:2009mn,Buckley:2009in, Feng:2009hw,  Feng:2010zp,Tulin:2012wi,Tulin:2013teo} but generally require a numerical solution. We use the code \texttt{CLASSICS}~\cite{Colquhoun:2020adl} to calculate the viscosity cross section numerically. Examples of the velocity dependence are shown in Fig.~\ref{fig:sig_V}.

Astrophysical observations indirectly probe the DM--DM self-interaction cross section in different environments with different characteristic DM velocities~\cite{Kaplinghat:2015aga,Tulin:2017ara,Bullock:2017xww,Sagunski:2020spe}. In dwarf galaxies, we have $v_\text{DM}\sim 10$--$\unit[100]{km/s}$ and preferred values for the cross section over DM mass of order $\sigma/m \sim \unit[1]{cm^2/g}$ to solve the core--cusp problem; in clusters, on the other hand, DM velocities are of the order of $v_\text{DM}\sim \unit[1000]{km/s}$ and self-interaction cross sections should be around or below $\sigma/m \sim \unit[0.1]{cm^2/g}$. These cross sections are significantly larger than the required annihilation cross sections for thermal freeze-out and hence non-trivial to obtain.
As can be seen in Fig.~\ref{fig:sig_V}, our model can indeed generate these large cross sections, at least for light $Z'$. In the thermal relic case, $\sigma/m \sim \unit[1]{cm^2/g}$ can be achieved in dwarf galaxies, but the cross sections are significantly suppressed on cluster scales. While still faring better than standard cold-DM models, the thermal relic case might therefore only provide a partial resolution to the core--cusp problem given the small preference of sizable self-interactions even at cluster scales~\cite{Sagunski:2020spe}.
Using instead \emph{asymmetric} DM -- where the coupling $q g'$ is allowed/required to be larger -- can easily provide a perfect fit for self-interactions both at dwarf and cluster scales.

\begin{figure}[tb]
	\includegraphics[width=0.52\textwidth]{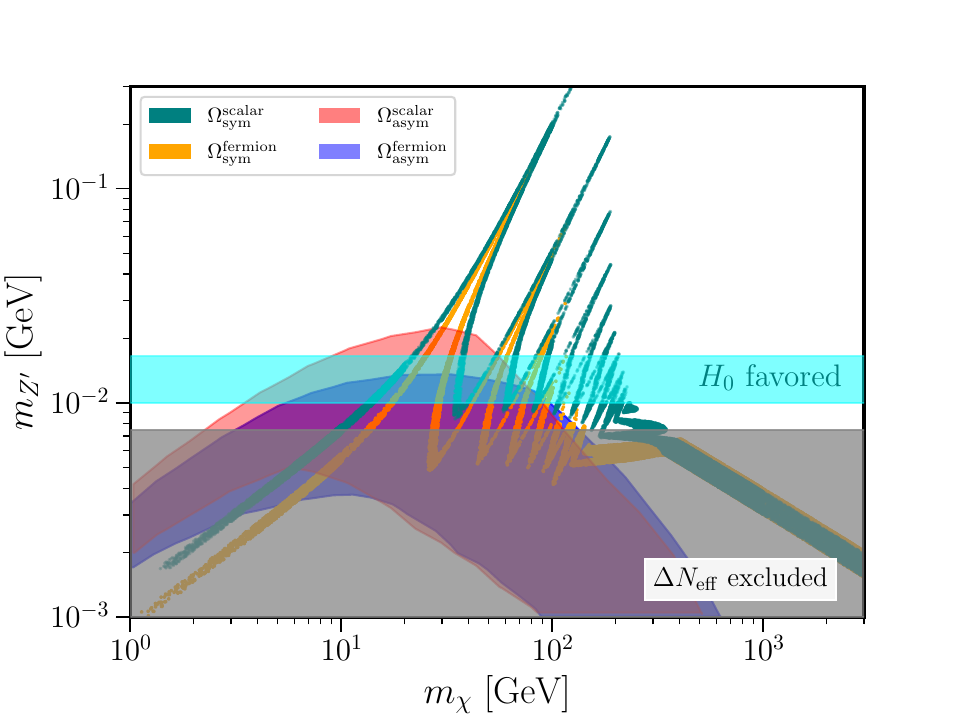}      
	\caption{Allowed parameter space of scalar (green) and fermionic Dirac (orange) DM mass $m_\chi$ and gauge boson mass $m_{Z'}$ that satisfies velocity-dependent self-interaction rate $\sigma/m \simeq \unit[1]{cm^2/g}$ ($< \unit[0.1]{cm^2/g}$) on dwarf-galaxy (cluster-galaxy) scales with DM velocity fixed at $v_{\rm dwarf} = \unit[10]{km/s}$ and $v_{\rm cluster} = \unit[1000]{km/s}$. Both of the data sets satisfy relic abundance requirement $\Omega h^2 \simeq 0.12$~\cite{Planck:2018vyg}. Red (blue) band corresponds to the allowed parameter space for asymmetric scalar (fermionic) DM (with $\Omega_\text{sym} h^2 \simeq 0.12/100$) with the required $\sigma/m$ and velocity to incorporate both dwarf and cluster galaxies data, as depicted in Fig.~\ref{fig:sig_V}. The purple region is the overlap of the red and blue bands. 
}
	\label{fig:DMself}
\end{figure}

Fixing the relic abundance to $\Omega_\text{sym} h^2 = 0.12$ (for thermal relic DM) or $\Omega_\text{sym} h^2 = 0.12/100$ (for asymmetric DM), we can scan over the two masses $m_\chi$ and $m_{Z'}$, demanding the DM self-interactions to solve the core--cusp problem, shown in Fig.~\ref{fig:DMself}. In the thermal relic case, the self-interactions are an average of the attractive $\bar{\chi}\chi \to  \bar{\chi}\chi$ and the repulsive $\chi\chi\to \chi\chi$ channel, which, in particular, results in a spiky behavior due to bound-state formation. In the asymmetric case, only the repulsive channel survives, which does not feature such bound states. In both cases we find an \emph{upper} limit on the $Z'$ mass that falls precisely in the region in which the $Z'$ can explain the $(g-2)_\mu$ anomaly (see Fig.~\ref{fig:Lmu-Ltau_limits}). Turning this around, the $(g-2)_\mu$-motivated region is exactly where we need the $Z'$ to be in order to give large DM self-interactions, a non-trivial coincidence that makes the $m_{Z'}\sim \unit[10]{MeV}$ region extremely special.
Even increasing $q g'$ up to its perturbativity bound of $\sqrt{4\pi}$ -- and thereby decreasing the symmetric component $\Omega_\text{sym}$ further -- only increases the $m_{Z'}$ upper bound to $\simeq \unit[120]{MeV}$.
For symmetric DM, one can push $m_{Z'}$ up to $\simeq\unit[700]{MeV}$ along the bound-state spikes in Fig.~\ref{fig:DMself} while still partially solving the core--cusp problem. 
Overall, resolving the small-scale structure problems unavoidably forces the $Z'$ to be light, almost precisely in the region where it can still explain $(g-2)_\mu$.

$L_\mu-L_\tau$ DM can give the correct abundance and resolve the small-scale structure problems with a light $Z'$ in the $(g-2)_\mu$-preferred region. Additional constraints on the model arise from indirect detection, which, in the symmetric-DM case, probes DM--anti-DM annihilations into SM particles. Since $q\gg 1$ in the region of interest, $\chi\bar{\chi}\to Z'Z'$ continues to dominate the annihilation, which is then followed by $Z'\to$\,neutrinos in the $Z'$ mass range of interest for $(g-2)_\mu$. This generates a polynomial spectral feature in the neutrino flux~\cite{Garcia-Cely:2016pse} that is out of reach even at next-generation neutrino detectors~\cite{Asai:2020qlp} for light DM. At the upper end of our DM mass range, Sommerfeld enhancement could lead to symmetric-DM annihilation rates in reach of KM3NeT~\cite{BasegmezDuPree:2021fpo}.
Similarly, constraints from energy injection into the cosmic microwave background are satisfied due to the dominant DM annihilation into neutrinos~\cite{Bringmann:2016din,Hambye:2019tjt}.
Our model is hence safe from indirect-detection constraints, at least in the region of interest. \emph{Direct} detection is absent at tree level since the $Z'$ only couples to second- and third-generation leptons; a non-zero kinetic mixing angle would enable DM scattering on nucleons~\cite{Garani:2019fpa}, but without any relation to the physics discussed so far.
As pointed out in Ref.~\cite{Garani:2019fpa}, the large number of muons in neutron stars would allow them to capture our muonphilic DM very efficiently and, in the process, heat up the neutron star. This is potentially observable with infrared telescopes such as the recently launched James Webb Space Telescope~\cite{Baryakhtar:2017dbj}.

\section{Conclusion}
\label{sec:conclusion}

The Standard Model of particle physics is currently facing numerous issues, ranging from the well-established neutrino mass and dark matter to lepton-flavor anomalies such as $(g-2)_\mu$, $R(K)$, and $R(D)$. In this article we have presented a simple model based on a gauged $U(1)_{L_\mu-L_\tau}$ symmetry extended by some leptoquarks that efficiently resolves all these issues and furthermore leads to velocity-dependent dark-matter self-interactions that can ameliorate current small-scale structure-formation discrepancies. For $Z'$ masses in the $\unit[10]{MeV}$ region, the Hubble tension can be ameliorated as well.
While other $U(1)'$ extensions can and have been used in connection to these issues, $L_\mu-L_\tau$ is unique in its ability to address all of them simultaneously, not least because all flavor anomalies reside in the muon and tauon sector.

\bibliographystyle{utcaps_mod}
\bibliography{BIB}

\end{document}